\documentstyle[preprint,prd,aps]{revtex}
\def\r2{\sqrt 2}
\def\delt{\partial/\partial t}
\def\PL{{1-\gamma_5\over 2}}
\def\PR{{1+\gamma_5\over 2}}
\def\v#1{v_#1}
\def\tb{\tan\beta}
\def\KK{$K^0$-$\bar{K^0}$}
\def\g{\lambda}
\def\gp{\lambda_{(+)}}
\def\gm{\lambda_{(-)}}
\def\agp{\bar\lambda_{(+)}}
\def\agm{\bar\lambda_{(-)}}
\def\la#1{\lambda_#1}
\def\lam{\lambda^-}
\def\h{\psi}
\def\hp{\psi_{(+)}}
\def\hm{\psi_{(-)}}
\def\ahp{\bar\psi_{(+)}}
\def\ahm{\bar\psi_{(-)}}
\def\ps#1{\psi_#1}
\def\psm{i\psi^-_1}
\def\psp{(i{\psi^+_2})^c}
\def\m#1{{\tilde m}_#1}
\def\mH{m_H}
\def\c{\omega _i}
\def\cp{\omega _{i(+)}}
\def\cm{\omega _{i(-)}}
\def\acp{\bar\omega _{i(+)}}
\def\acm{\bar\omega _{i(-)}}
\def\mw#1{{\tilde m}_{\omega #1}}
\def\mwi{{\tilde m}_{\omega i}}
\def\plam{p_\lambda}
\def\ppsi{p_\psi}
\def\p#1{p_#1}
\def\dW{\delta_W}
\def\vW{v_W}
\def\vW2{v_W^2}

\begin{document}
\draft
\preprint{OCHA-PP-87}
\title{ 
Electroweak baryogenesis from chargino transport in 
the supersymmetric model
}
\author{
Mayumi Aoki, Noriyuki Oshimo, and Akio Sugamoto
}
\address{
Department of Physics, Ochanomizu University  \\
Otsuka 2-1-1, Bunkyo-ku, Tokyo 112, Japan  \\
}
\date{\today}
\maketitle
\begin{abstract}
     We study the baryon asymmetry of the universe in  
the supersymmetric standard model (SSM).  
At the electroweak phase transition, the fermionic 
partners of the charged SU(2) gauge bosons and Higgs 
bosons are reflected from or transmitted to the bubble wall  
of the broken phase.  
Owing to a physical complex phase in their mass matrix, 
these reflections and transmissions 
have asymmetries between CP conjugate processes.  
Equilibrium conditions in the symmetric phase 
are then shifted to favor a 
non-vanishing value for the baryon number density, which is 
realized through 
electroweak anomaly.  We show that the resultant 
ratio of baryon number to entropy is consistent   
with its present observed value within  
reasonable ranges of SSM parameters, provided that 
the CP-violating phase intrinsic in the SSM 
is not much suppressed.  
The compatibility with the constraints on the parameters 
from the electric dipole moment of the neutron is also 
discussed.       
\end{abstract}

\pacs{11.30.Er, 11.30.Fs, 12.15.Ji, 12.60.Jv, 98.80.Cq}

\narrowtext

\section{Introduction}
\label{sec1}

     The astronomical observations indicate that there 
exist more baryons than antibaryons in our universe.  
This baryon asymmetry of the universe may be 
understood within the framework of physics at the 
electroweak scale \cite{kuzmin}, 
since all the necessary ingredients for 
baryogenesis could be available there \cite{ewbrev}.   Although 
the standard model (SM) does not account for the asymmetry 
quantitatively, certain extensions of the SM would be able  
to overcome those difficulties which the SM encounters.   
This possibility could give some hints for 
physics beyond the SM.  In  particular, 
it turned out that CP violation in the SM arising from 
the Kobayashi-Maskawa phase does not  
lead to an enough amount of asymmetry between the 
numbers of baryons and antibaryons.   
The baryon asymmetry could be a unique phenomenon ever 
found, other than the \KK\ system, which enables us to 
study CP violation.    

     In this paper, we discuss the possibility of baryogenesis 
at the electroweak phase transition  
in the supersymmetric standard model (SSM). 
This model \cite{ssmrev}, which is one of the most plausible 
extensions of the SM from the viewpoint of physics at the 
electroweak scale, contains new sources of CP violation \cite{cp}  
wanted for the baryogenesis.  In addition, 
the electroweak phase transition may be strongly first order 
in the SSM, which is also necessary for the 
baryogenesis though not compatible with the Higgs 
boson mass in the SM.  
Indeed, the constraints on the Higgs boson masses  
from this requirement could be 
relaxed \cite{ewpt} compared to the SM, 
owing to the richness of Higgs fields, 
especially by adding a gauge singlet field.  
It would be thus of great importance to study 
whether the electroweak baryogenesis is viable in   
the SSM.  

     The baryogenesis at the electroweak 
phase transition could occur within  
or outside the bubble wall of the broken phase, 
where baryon number violation  
by electroweak anomaly is 
not suppressed.   
We consider the baryon number generation outside 
the wall, 
which has been suggested to give an enough amount of 
asymmetry in the SSM \cite{nelson1,nelson2}.   
For this process the baryon number is well  
estimated through the  
charge transport mechanism \cite{nelson1,ctm}.  
The mediators on which our study is focused are 
charginos, which consist of the 
fermionic partners of the charged 
SU(2) gauge bosons and Higgs bosons and have a 
mass matrix with a physical complex phase.    
Since these particles couple to the Higgs bosons 
by the SU(2) gauge interaction, their scatterings 
at the wall are not so weak as the leptons, 
while their asymmetries in the symmetric phase 
are maintained longer than the quarks.     
It will be shown that the baryon 
asymmetry of the universe can be explained in the SSM 
within the reasonable values for its model parameters.  
However, the allowed range for the new CP-violating phase is not 
so wide as estimated before \cite{nelson1,nelson2}, 
so that the baryogenesis 
could give nontrivial constraints on the SSM.  
Although the baryogenesis within the wall might also be possible 
in the SSM \cite{spon},     
it has been generally shown that an enough 
amount of baryon asymmetry 
cannot be produced within the wall only \cite{dine}.  
 
     In Sec. \ref{sec2} we briefly review the source of 
CP violation in the chargino sector.  In Sec. \ref{sec3} 
we discuss CP asymmetries in the reflection and 
transmission rates for the charginos at the bubble 
wall.  The procedure for computing these rates, which 
gives accurate results, is presented explicitly.  
In Sec. \ref{sec4} we calculate the ratio of baryon number to 
entropy following the charge transport mechanism.  
The dependences of the ratio on various parameters are also analyzed.  
Summary is given in Sec. \ref{sec5}.  

\section{Supersymmetric model} 
\label{sec2}

     A new source of CP violation in the SSM comes from 
the mass matrices of the SU(2)$\times$U(1) gauginos 
and Higgsinos.   The mass terms for the charged gauginos $\lam$ 
and Higgsinos $\psm$, $\psp$ are given by 
\begin{eqnarray}
 {\cal L}&=&-(\overline{\lam} \quad \overline{\psp})M^-\PL
    \left(\matrix{\lam  \cr
                  \psm }  \right)
              +{\rm h.c.},   \\
\label{1} 
    M^- &=& \left(\matrix{\m2 & -g\v1^*/\r2 \cr
                -g\v2^*/\r2 & \mH}        \right),    
\label{2} 
\end{eqnarray}
where $\m2$ denotes the mass parameter for the SU(2) 
gauginos arising from the supersymmetry soft-breaking 
term; $\mH$ denotes the mass parameter for the 
Higgsinos from the bilinear term of Higgs superfields 
in superpotential; and $\v1$ and $\v2$ are respectively 
the vacuum expectation values of the Higgs bosons 
with U(1) hypercharges $-1/2$ and $1/2$.  
The mass matrix (\ref{2}) is diagonalized by unitary 
matrices $C_R$ and $C_L$ as
\begin{equation}
      C_R^\dagger M^-C_L = {\rm diag}(\mw1, \mw2) \quad 
                       (\mw1 <\mw2 ),     
\label{3}
\end{equation}
giving the mass eigenstates for the charginos $\c$.  

     In general, the parameters $\v1$, $\v2$, $\m2$, and $\mH$ 
in the mass matrix (\ref{2}) have complex values.  
Although there is some freedom of defining phases for the particle fields, 
if the SU(2)$\times$U(1) gauge symmetry is spontaneously  
broken and thus $\v1$ and $\v2$ have non-vanishing values, 
all the complex phases cannot be rotated 
away.  The redefinitions of the fields make it possible 
without loss of generality to take $\m2$, $\v1$, and $\v2$ 
for real and positive.  Then $\mH$ cannot be made real.  
Therefore, there is one physical complex phase in the 
mass matrix for the charginos, which we express as 
\begin{equation}
\mH=|\mH|\exp(i\theta).  
\label{4}
\end{equation}
Owing to this complex phase, CP invariance is broken in the 
interactions for the charginos.  
Similarly the mass matrix for the neutral gauginos 
and Higgsinos contains the CP-violating phase.   

\section{CP asymmetry} 
\label{sec3}

     At the electroweak phase transition of the universe, 
if it is first order, 
bubbles of the broken phase nucleate in 
the SU(2)$\times$U(1) symmetric phase.  
In the symmetric phase the gauginos and the Higgsinos 
are in mass eigenstates themselves.  On the other hand, they are mixed 
to form mass eigenstates in the 
wall and in the broken phase, owing to non-vanishing 
vacuum expectation values of the Higgs bosons.  
Consequently, the gauginos incident on the wall from  
the symmetric phase can be reflected to become  
Higgsinos, and vice versa.  The charginos from the broken phase 
can be transmitted to the symmetric phase and become gauginos 
or Higgsinos.  In these processes CP violation  
makes a difference in reflection or 
transmission probability between a  
particle state with a definite helicity and its 
CP-conjugate state.  The induced CP asymmetries shift equilibrium 
conditions in the symmetric phase for non-vanishing baryon number.   

     The reflection and transmission rates at the wall are   
obtained by solving the Dirac equations for the charginos.   
In the rest frame of the wall the Dirac equations are given by 
\begin{equation}
 \left(\matrix{-i\delt &  \m2  & 0 & -g\v1^*/\r2 \cr
                         -\m2^*  & i\delt & g\v2/\r2 &   0 \cr
                          0  & -g\v2^*/\r2 &   -i\delt & \mH \cr
                      g\v1/\r2 &  0 & -\mH^* &   i\delt}       \right) 
  \left(\matrix{\la1 \cr
                         \la3 \cr
                         \ps1 \cr
                         \ps3 \cr}    \right) 
                   =i\frac{\partial}{\partial z}   
  \left(\matrix{\la1 \cr
                         \la3 \cr
                         \ps1 \cr
                         \ps3 \cr}    \right), 
\label{5}
\end{equation}
\begin{equation}
 \left(\matrix{-i\delt &  \m2^*  & 0 & -g\v2/\r2 \cr
                         -\m2  & i\delt & g\v1^*/\r2 &   0 \cr
                          0  & -g\v1/\r2 &   -i\delt & \mH^* \cr
                      g\v2^*/\r2 &  0 & -\mH &   i\delt}      \right) 
  \left(\matrix{\la4 \cr
                         \la2 \cr
                         \ps4 \cr
                         \ps2 \cr}   \right)
                  =i\frac{\partial}{\partial z}
   \left(\matrix{\la4 \cr
                         \la2 \cr
                         \ps4 \cr
                         \ps2 \cr}   \right),                   
\label{6}
\end{equation}
where the wall is taken to be parallel to the $xy$-plane and 
perpendicular to the velocity of the particles.  
The components of the Dirac fields are expressed as 
\begin {equation}
     \lam\equiv\left(\matrix{\la1 \cr
                             \la2 \cr
                             \la3 \cr
                             \la4 \cr}  \right),   \quad         
  \PL\psm+\PR\psp\equiv\left(\matrix{\ps1 \cr
                                     \ps2 \cr
                                     \ps3 \cr
                                     \ps4 \cr}  \right).  
\label{7}
\end{equation}
We have adopted the chiral representation for the Dirac $\gamma$ 
matrices.        
In the symmetric phase the vacuum expectation values $\v1$ and 
$\v2$ vanish, while in the broken phase they are related to the  
$W$-boson mass as $M_W=(g/2)\sqrt{|\v1|^2+|\v2|^2}$.  
The vacuum expectation values vary along the $z$-axis in 
the wall.  For simplicity, we assume that the wall is situated from 
$z=0$ to $z=2\dW$ and the $z$-dependences of $\v1$ 
and $\v2$ are given by  
\begin{eqnarray}
   \sqrt{|\v1|^2+|\v2|^2}&=&
           \frac{M_W}{g}\{1+\tanh(\frac{z}{\dW}-1)\pi\}, \nonumber \\
   \frac{\v2}{\v1}&=&\tb,  
\label{8}
\end{eqnarray}
where the ratio $\v2/\v1$ is taken to be constant and equal 
to its value $\tb$ in the broken phase.  
The symmetric phase is in the region $z<0$.  

     As a prototype for the calculation of the reflection and 
transmission rates, we consider the case that a gaugino 
with a positive helicity and energy $E$ enters 
from the symmetric phase.  
Then the reflected particle (gaugino, Higgsino) has a negative helicity 
and the transmitted particle (charginos) has a positive helicity 
by angular momentum conservation.     
The boundary condition at $z=0$ becomes 
\begin{equation}
\left(\matrix{\la1 \cr
                         \la3 \cr
                         \ps1 \cr
                         \ps3 \cr}    \right)=
			 \{X_1(0)+AX_2(0)+BX_3(0)\}\exp(-iEt),   
\label{9}
\end{equation}
\begin{eqnarray}
       X_1(0)&=&\sqrt{\frac{E+\plam}{2\plam}}
\left(\matrix{1 \cr
              \m2/(E+\plam) \cr 
              0 \cr
              0 \cr}              \right),   \quad 
      X_2(0)=\sqrt{\frac{E+\plam}{2\plam}}
\left(\matrix{ \m2/(E+\plam) \cr
                         1 \cr
                         0 \cr
                         0 \cr}    \right),  \nonumber \\     
      X_3(0)&=&\exp(-i\theta)\sqrt{\frac{E+\ppsi}{2\ppsi}}             
\left(\matrix{0 \cr
              0 \cr
              \mH/(E+\ppsi) \cr
              1 \cr}           \right),     
\label{10}
\end{eqnarray}
where $\plam$ and $\ppsi$ stand for the absolute values 
of the momenta for the gaugino and the Higgsino.  
The incident gaugino, 
the reflected gaugino, and the reflected Higgsino correspond to 
$X_1$, $X_2$, and $X_3$ , respectively.  
The boundary condition at $z=2\dW$ becomes 
\begin{equation}
 \left(\matrix{\la1 \cr
                         \la3 \cr
                         \ps1 \cr
                         \ps3 \cr}    \right)
           =\{CY_1(2\dW)+DY_2(2\dW)\}\exp(-iEt),   
\label{11}
\end{equation}   
\begin{eqnarray}
       Y_1(2\dW)&=&\sqrt{\frac{E+\p1}{2\p1}}
\left(\matrix{C_{R11} \cr
              C_{L11}\mw1/(E+\p1) \cr 
              C_{R21} \cr
              C_{L21}\mw1/(E+\p1) \cr}  \right)\exp(i2\p1\dW), 
	      \nonumber \\
      Y_2(2\dW)&=&\sqrt{\frac{E+\p2}{2\p2}}
\left(\matrix{C_{R12} \cr
              C_{L12}\mw2/(E+\p2) \cr
              C_{R22} \cr
              C_{L22}\mw2/(E+\p2) \cr}  \right)\exp(i2\p2\dW),
\label{12}
\end{eqnarray} 
where $\p1$ and $\p2$ stand for the absolute values 
of the momenta for the two charginos.  
The lighter and heavier 
charginos correspond to $Y_1$ and $Y_2$, respectively.  
For the wave functions  
$X_1$, $X_2$, and $X_3$ given at $z=0$,  
those at $z=2\dW$ are obtained by numerically 
solving the differential equation (\ref{5}).  
The reflection and transmission amplitudes 
$A$, $B$, $C$, and $D$ then satisfy the simultaneous 
equation 
\begin{equation}
     X_1(2\dW)+AX_2(2\dW)+BX_3(2\dW)=CY_1(2\dW)+DY_2(2\dW), 
\label{13}
\end{equation}
which is solved algebraically.  The reliability of these 
numerical calculations may be checked by the sum  
$|A|^2+|B|^2+|C|^2+|D|^2$, which is in excellent agreement with 
unity in our results.  

      We calculate  
the asymmetries of the transition rates between CP 
conjugate processes    
\begin{eqnarray}
       A_\g&=&R(\gp\rightarrow\hm)+R(\gm\rightarrow\hp)
            -R(\agm\rightarrow\ahp)-R(\agp\rightarrow\ahm),  
\nonumber   \\
      A_{\c}&=&R(\cp\rightarrow\hp)+R(\cm\rightarrow\hm)
                 -R(\acm\rightarrow\ahm)-R(\acp\rightarrow\ahp), 
\label{14}
\end{eqnarray}
where $\g$ and $\h$ denote the gaugino and the Higgsino in the 
symmetric phase, respectively, and $\c$ the chargino in the 
broken phase.  
The subscripts $(+)$ and $(-)$ refer to a positive 
and a negative helicities, respectively.  
If CP is not violated, $A_\g$ and $A_{\c}$ vanish.  
In Fig. \ref{fig1} the absolute values of  
these CP asymmetries are shown for (a) $A_\g$, 
(b) $A_{\omega_1}$, and (c) $A_{\omega_2}$ 
as functions of the particle energy $E$, 
taking $\theta=\pi/4$, 
$\tb=2$, $\m2=|\mH|=200$ GeV, and $\dW=1/200$ (GeV)$^{-1}$. 
The sign of $A_\g$ is negative for $E<\mw2$ and positive for 
$E>\mw2$, and those of $A_{\omega_1}$ and $A_{\omega_2}$ 
are positive and negative, respectively.  
The sum of $A_\g$, $A_{\omega_1}$, and $A_{\omega_2}$ 
vanishes by CPT invariance.  
The CP asymmetry $A_\g$ has values of order of 0.1
for the energy range in slightly excess of the particle mass 
and has much smaller values in the other energy range.  
Since the mass of the lighter chargino becomes smaller than 
that of the gaugino, the gaugino is transmitted to the broken 
phase at a large rate.  This makes the magnitude of  
$A_\g$ rather small  
in spite of an unsuppressed value for the CP-violating phase.  

\section{Baryon asymmetry}
\label{sec4}

     The CP asymmetries of the reflection and  
transmission rates for the charginos lead to a bias on 
equilibrium conditions favoring baryon asymmetry in the 
symmetric phase.  
The free energy 
is then minimized at a non-vanishing baryon number, 
to which the initial equilibrium state with no baryon 
asymmetry approaches through  electroweak anomaly.   
A simple procedure for relating the CP asymmetries  
with the bias \cite{nelson1} is to introduce chemical 
potentials for conserved and approximately conserved quantum 
numbers and set these quantum numbers for zero.  
This makes the chemical potential of baryon number 
given by a hypercharge density, which is induced by 
the CP asymmetries.  Although this procedure may not 
be completely correct, it would be able to 
provide a rough estimate for the bias.  

     Since the hypercharges of the gauginos 
and the Higgsinos are  $0$ and $-1/2$, respectively, 
the bubble wall emits a net flux of hypercharge by CP violation. 
The transitions which cause a change of hypercharge 
in the symmetric phase are 
(i) $\g\rightarrow\h$,  
(ii) $\h\rightarrow\g$, $\h\rightarrow\c$,   
(iii) $\c\rightarrow\h$, 
and their CP-conjugate transitions.  
The sum of the probabilities for the transitions in  
the reaction (ii) is however the same 
as that for their CP conjugate transitions by CPT invariance, so that 
a net hypercharge flux can be induced through 
the reactions (i) and (iii).  
 
     The hypercharge flux is  
calculated by convoluting the transition rates with 
the thermal flux of incoming particles.  In the thermal frame 
at temperature $T$, the 
incoming flux from the symmetric phase and 
that from the broken phase, $f_s$ and $f_b$, are given by 
\begin{eqnarray}
  f_s&=&\int\int\int_{D_s}\frac{d^3{\bf p}}{(2\pi)^3}
                   (\frac{p_z}{E_T}+v_W)[\exp(\frac{E_T}{T}+1)]^{-1} 
           \quad  (D_s:  \frac{p_z}{E_T}+v_W>0),       \nonumber \\
f_b&=&-\int\int\int_{D_b}\frac{d^3{\bf p}}{(2\pi)^3}
                   (\frac{p_z}{E_T}+v_W)[\exp(\frac{E_T}{T}+1)]^{-1} 
           \quad  (D_b:  \frac{p_z}{E_T}+v_W<0),  
\label{15}
\end{eqnarray}
where $E_T$ and $p_z$ represent the total energy and 
the $z$-component of the momentum for the particle.  
The wall is taken to be perpendicular to the $z$-axis 
and moving with the velocity $-v_W (v_W>0)$.  
The net hypercharge flux therefore becomes  
\begin{eqnarray}
  F_Y&=&F_\g+\sum_{i=1}^2F_{\c},   
\label{16}    \\
  F_\g&=&-\frac{1}{2}\frac{(1-\vW2)T}{(2\pi)^2}
    \int_{\m2}^\infty dE
      E\log[1+\exp\left(
      -\frac{E-v_W\sqrt{E^2-\m2^2}}{T\sqrt{1-\vW2}}
		  \right)]
               A_\g,            \nonumber \\
F_{\c}&=&-\frac{1}{2}\frac{(1-\vW2)T}{(2\pi)^2}
    \int_{\mwi}^\infty dE
      E\log[1+\exp\left(
	    -\frac{E+v_W\sqrt{E^2-\mwi^2}}{T\sqrt{1-\vW2}}
		   \right)]
               A_{\c},             \nonumber 
\end{eqnarray}
where $E$ represents $\sqrt{p_\perp^2+m^2}$, $p_\perp$ and $m$ 
denoting the component of the momentum perpendicular to 
the wall in the wall rest frame and the mass, respectively, for 
the relevant particle.     

     Assuming detailed balance for the transitions among 
the states of different baryon numbers, the rate equation of the 
baryon number density $\rho_B$ is given by \cite{dine2} 
\begin{equation}
   \frac{d\rho_B}{dt}=-\frac{\Gamma}{T}\mu_B, 
\label{17}
\end{equation}
where $\Gamma$ denotes the rate per unit 
time and unit volume for the transition between the neighboring 
states different by unity in baryon number, and $\mu_B$ stands for the 
chemical potential of baryon number which represents 
a bias on the fluctuation of baryon number.  
In the symmetric phase the rate $\Gamma$ is estimated as 
\begin{equation} 
   \Gamma=3\kappa(\alpha_WT)^4 ,  
\label{18}
\end{equation}
where $\kappa$ is of order of unity \cite{ambjorn}.  

     The chemical potential $\mu_B$ may be 
related to the hypercharge density through equilibrium 
conditions \cite{ewbrev,nelson1}.  In the symmetric phase   
the gauge interactions and the $t$-quark Yukawa 
interactions are considered to be in chemical equilibrium.  
We also take that the self interactions of the Higgs bosons, 
the Higgsinos, and the gauginos are in equilibrium, respectively.  
Among the supersymmetric particles, the squarks, sleptons, 
and gluinos are assumed to be heavy enough.  
Before the net hypercharge flux is emitted from the wall, 
the baryon number and the lepton number vanish.  
We thus impose for equilibrium conditions 
vanishing values on the baryon  
and the lepton number densities in each generation, and  
on the number densities of the 
right-handed quarks and leptons except the $t$-quark. 
These constraints lead to the chemical potential for baryon number  
given by 
\begin{equation}
     \mu_B=-\frac{2\rho_Y}{7T^2},   
\label{19}
\end{equation}
where $\rho_Y$ stands for the hypercharge density.  
The net hypercharge flux into the symmetric phase 
induces a net hypercharge density, which makes  
$\mu_B$ non-vanishing.  
 
     We now estimate the baryon number density $\rho_B$ 
from Eqs. (\ref{17}), (\ref{18}), and (\ref{19}) as   
\begin{eqnarray} 
   \rho_B&=&\frac{2\Gamma}{7T^3}
   \int_{-\infty}^{z/v_W}dt\rho_Y(z-v_Wt)
   =\frac{2\Gamma}{7T^3v_W}\int_0^\infty dz\rho_Y(z)
   \nonumber \\
   &\approx&\frac{2\Gamma F_Y\tau_T}{7T^3v_W}, 
\end{eqnarray}
where $\tau_T$ is the time which carriers of the hypercharge 
flux spend in the symmetric phase.  This transport time $\tau_T$ 
may be approximated by the mean free time of the carriers. 
The rough estimate for $\tau_T$ gives a value 
of order of $10^2/T-10^3/T$ for the leptons \cite{joyce},   
which would also be applicable to the gauginos and the 
Higgsinos.  
The ratio of the baryon number to entropy is given by 
\begin{equation}
  \frac{\rho_B}{s}=\frac{135\kappa\alpha_W^4F_Y\tau_T}
                                       {7\pi^2g_*v_WT^2},
\end{equation}
where $g_*$ represents the relativistic degrees of freedom 
for the particles.  For definiteness, we take $g_*=124.75$, 
where SU(2)$\times$U(1) gauginos and Higgsinos are 
taken into account as well as gauge bosons, Higgs bosons, 
quarks, and leptons.    

     We show the ratio of baryon number to entropy 
in Fig. \ref{fig2} as a function of the mass parameter 
$\m2 (=|\mH|)$ for $\tb=2$ and 
(a) $\theta=\pi/4$ and (b) $\theta=0.1$.  
For simplicity, we have taken the same value for 
$\m2$ and $|\mH|$.  
In the mass range where curves are 
not drawn, the lighter chargino has a mass 
smaller than 45 GeV, which is ruled out by 
experiments \cite{pdg}.  
The temperature is taken for $T=200$ GeV.   
The wall velocity $v_W$ and the wall 
width $\dW$ have been estimated in the SM as 
$v_W=0.1-1$ and $\dW\sim 10/T$ \cite{turok}, 
although there are large uncertainties
and model dependences.  
If the phase transition is strongly first order, the wall 
width generally becomes thinner.  
We take four sets of 
values for $v_W$ and $\dW$ listed in Table \ref{tab1},  
which correspond to four curves (i.a)-(ii.b).  
For definiteness we set the transport time for $\tau_T=100/T$.  
The resultant ratio is around $10^{-10}$ for $\theta=\pi/4$ and 
$10^{-11}$ for $\theta=0.1$, which are consistent with 
the present observed value 
$\rho_B/s=(2-9)\times 10^{-11}$ \cite{pdg}. 
Except the CP-violating phase $\theta$, 
the ratio does not vary much with the SSM parameters 
$\m2$, $|\mH|$, and $\tb$.  
In Fig. \ref{fig3} the ratio of baryon number to entropy 
is shown as a function of $\theta$ 
for (i) $T=100$ GeV and (ii) $T=200$ GeV, taking 
$\tb=2$, $\m2=|\mH|=200$ GeV, $v_W=0.6$, and $\dW=1/T$.   
If the CP-violating phase is  
of order of $0.01$, only a large value for $\tau_T$ 
of order of $10^{3}/T$ barely gives a compatible 
value.  

     In the SSM the CP-violating phase $\theta$ 
gives large contributions to the electric 
dipole moments (EDMs) of the neutron and the electron at 
one-loop level through the diagrams mediated by 
the charginos \cite{edm}.  In Fig. \ref{fig4} 
the neutron EDM is shown as a function of $\m2 (=|\mH|)$ 
for $\tb=2$. Four curves correspond to  
four sets of values for $\theta$ and the mass for the $u$- and 
$d$-squarks, 
which are listed in Table \ref{tab2}.  
The experimental upper bound on the magnitude of the neutron EDM 
is about $10^{-25} e$cm \cite{pdg}.  
In order to satisfy this bound, 
the squark mass should be around or larger than 1 TeV 
for $\theta\sim 0.1$ and 3 TeV for $\theta\sim 1$.  
Similar constraints are obtained from the EDM of the 
electron.  

\section{Summary}
\label{sec5}

     We studied the electroweak baryogenesis 
mediated by the charginos in the SSM following 
the charge transport mechanism.    
The observed baryon asymmetry is                  
explained in reasonable ranges of the SSM parameters.  
However, the CP-violating phase cannot be so small as 
estimated in the literature.  
This is mainly because the gauginos or the Higgsinos 
incident on the wall from the symmetric phase can 
be transmitted to the broken phase, thus reducing 
the overall reflection rates in the wide region of 
the parameter space.  
Assuming a moderate value 
for the transport time, the phase should be at least of 
order of $0.1$. If this is the case, the squark masses 
are predicted to 
be around or larger than 1 TeV from the analysis 
of the EDM of the neutron.   

     The SSM has some ambiguities in the Higgs sector.  
The minimal version of the SSM contains two Higgs 
doublets and no singlet.  However, having a 
singlet field may well be motivated from various standpoints 
in particle physics, which could make the electroweak 
phase transition strongly first order.  
It should be noted that the chargino sector relevant to our 
discussions does not 
depend much on whether the SSM has a singlet or not,  
so that our analyses are mostly applicable to both cases.  
The baryon asymmetry could also be generated 
by the neutralinos which consist of neutral SU(2)$\times$U(1) 
gauginos and Higgsinos.  This neutralino contribution 
would be at most the same order of magnitude as the
chargino contribution.  

\acknowledgements
     One of the authors (N.O.) thanks A.E. Nelson for informing 
him of their work \cite{nelson1}.  
We thank T. Uesugi and K. Yamamoto for helpful discussions.  
This work is supported in part by the Grant-in-Aid for 
Scientific Research for the Ministry of Education, Science 
and Culture (No. 08640357).  

%

%
%
%
\begin{figure}
\caption{The CP asymmetries 
for the reflections and transmissions
at the bubble wall as functions of the particle energy:  
(a) $A_\g$, (b) $A_{\omega_1}$, (c) $A_{\omega_2}$.
The parameter values are taken for $\theta=\pi/4$, $\tb=2$, 
$\m2=|\mH|=200$ GeV, and $\dW=1/200$ (GeV)$^{-1}$. } 
\label{fig1}
\end{figure}         
\begin{figure}
\caption{The ratio of baryon number to entropy as a function of 
$\m2(=|\mH|)$:  
   (a) $\theta=\pi/4$, (b) $\theta=0.1$.  
        The values of $v_W$ and $\dW$ for curves (i.a)--(ii.b) 
         are given in Table \ \protect\ref{tab1}.
      The other parameters are taken for $\tb=2$ and $T=200$ GeV.  }
\label{fig2}
\end{figure}         
\begin{figure}
\caption{The ratio of baryon number to entropy 
as a function of $\theta$ for (i) $T=100$ GeV and 
(ii) $T=200$ GeV.  The parameters are taken for 
$\tb=2$, $\m2=|\mH|=200$ GeV, $v_W=0.6$, and $\dW=1/T$.  }
\label{fig3}
\end{figure}        
\begin{figure}
\caption{The electric dipole moment of the neutron 
   as a function of $\m2(=|\mH|)$.  
              The values of $\theta$ and the squark mass 
           for curves (i)--(iv) 
         are given in Table \ \protect\ref{tab2}.}
\label{fig4}
\end{figure}       
%
%
%
%
\begin{table}
\caption{The values of $v_W$ and $\dW$ for curves 
            (i.a)--(ii.b) in Fig.\ \protect\ref{fig2}.} 
\label{tab1}
\begin{tabular}{ccccc}
    & (i.a) & (i.b) & (ii.a) &(ii.b) \\
\tableline
 $v_W$  &  0.1 & 0.1 & 0.6 & 0.6 \\
 $\dW$  & 1/T  & 5/T  & 1/T & 5/T  \\
\end{tabular} 
\end{table}
\begin{table}
\caption{The values of $\theta$ and the squark mass for curves 
                (i)--(iv) in Fig.\ \protect\ref{fig4}.    }
\label{tab2}
\begin{tabular}{ccccc}
    & (i) & (ii) & (iii) &(iv) \\
\tableline
 $\theta$  & $\pi/4$ & $\pi/4$ & 0.1 & 0.1 \\
  Squark mass   (TeV) & 1  & 5  & 1 & 5  \\
\end{tabular} 
\end{table}

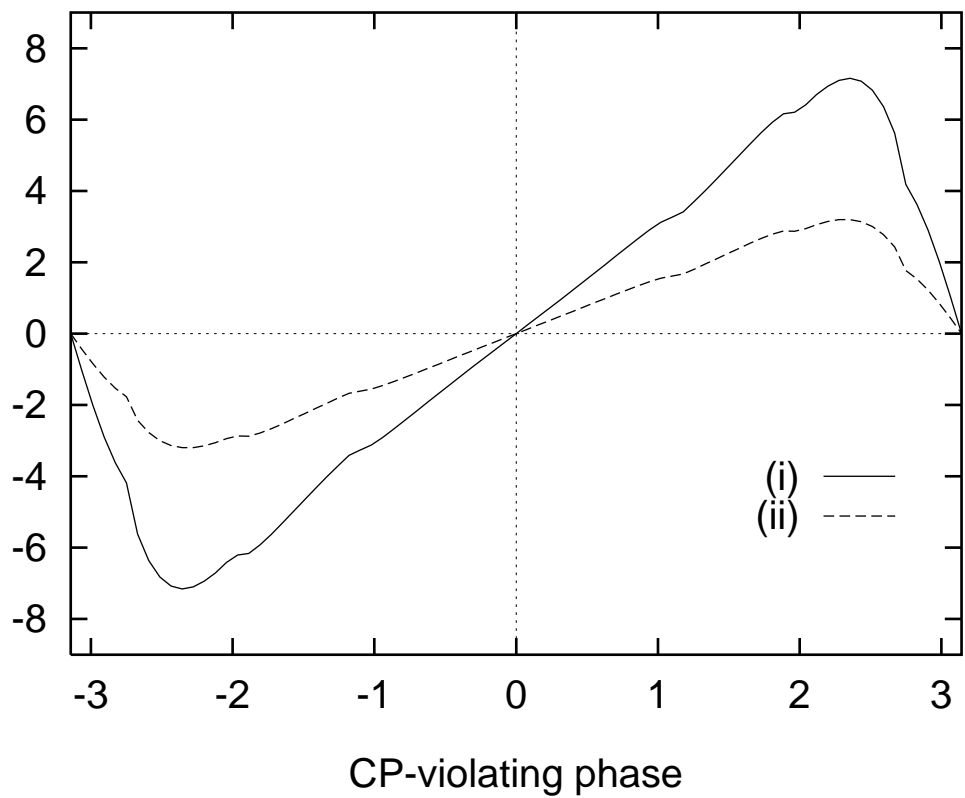
\end{document}